\documentclass[twocolumn,twoside,slac_two]{revtex4}
\usepackage{graphicx}
\usepackage{fancyhdr}
\newcommand{\lsp}{LS~I~+61$^{\circ}$303}
\newcommand{\lsi}{LS~I~+61$^{\circ}$303~}
\newcommand{\degr}{$^{\circ}$}

\begin{document}

\title{Core-shift and precession in the jet of \lsi}

\author{M. Massi}
\affiliation{Max-Planck-Institut f\"ur Radioastronomie, Auf dem H\"ugel 69, D-53121 Bonn, Germany}
\author{E. Ros}
\affiliation{Departament d'Astronomia i Astrof\'{\i}sica, Univ. de Val\`encia, E-46100 Burjassot,
Valencia, Spain}
\affiliation{Max-Planck-Institut f\"ur Radioastronomie, Auf dem H\"ugel 69, D-53121 Bonn, Germany}

\author{L. Zimmermann}
\affiliation{Max Planck Institut f\"ur Radioastronomie, Auf dem H\"ugel 69, D-53121 Bonn, Germany}

\begin{abstract}
\lsi is one of the few GeV- and TeV-emitting X-ray binaries with a prominent,
well-studied modulated radio and gamma-ray emission.
Changes in its  radio morphology  suggested in the past the hypothesis of a precessing microquasar.
In 2006, a set of VLBA observations performed all around the orbit and confirming the fast variation in morphology were not used to study the precession
because  the souce was explained in the context of the  pulsar model, the alternative model for this system.
 However, a recent  radio spectral index analysis over 6.7 years 
from Green Bank Interferometer data at 2.2 GHz and 8.3 GHz
has well  confirmed the predictions of the  microquasar scenario  in \lsp.
At the light of these results we reanalysed the set of VLBA observations that constitutes a unique tool to  determine the precession  period
and render a better understanding of the physical mechanism behind the precession.
We  improved the dynamic range of the images by a factor of four  using self-calibration,
and the self-calibrated maps reveal,
in six out of ten  images, a double-sided structure. 
The  double-sided structure  has variable position angle and 
switches at some epochs to a one-sided structure. These variations
  indicate a
scenario where the precessing jet, inducing variable Doppler boosting,  points close to our line of sight - a microblazar, the
galactic version of the extra-galactic blazars.
High energy observations of \lsi are consistent with the microblazar nature of this object.
Moreover, we suggest
in \lsi the first case of core shift effect observed in a microquasar.
Because of this effect, well known in  AGN, the cm-core of the jet  is rather
displaced from the system center. In \lsp, the cm-core  of the jet traces
a large ellipse,  7 times larger than the orbit, in  a period of about 28~d. Our
hypothesis is that this ellipse is the cross-section of the precession cone of the
jet at the distance of the 3.6 cm-core, and its period is the precession period.
\end{abstract}

\maketitle

\thispagestyle{fancy}


\section{Introduction}
\lsi is a high-mass X-ray binary (HMXB) where
  the compact object travels through the dense equatorial wind of  a rapidly rotating B0 Ve star.
  The nature of the compact object is still unknown due to    its
  large mass range, 1.4\,$M_{\odot} <  M    < 3.8\,M_{\odot}$, which  implies either  a neutron star or a black hole.

Two models have been proposed for this special HMXB, a strong and variable source  at all wavelenghts of
 the electromagnetic spectrum, from radio to TeV. One model assumes that the compact object is
 a non-accreting young pulsar whose
  relativistic wind strongly interacts with  the wind of  the Be star.
  The second model instead
   proposes a microquasar, that is,  an accreting object whose  steady jet, perpendicular to the accretion disk,
 is  occasionally travelled by shocks associated to transients.
   The peculiarity of \lsi is that, due to the eccentricity of the system,  two episodes of a large mass accretion rate may occur along the orbit and consequently two transients may occur per   orbital period (26.496 d) 
  (see references in [1]).

In 1993, VLBI observations of \lsi showed that the radio emission had a structure of milliarcsecond
(mas) size corresponding to a few AU at the distance of 2.0~kpc  \cite{massi93}.
However, the complex
morphology in successive VLBI observations
\cite{peracaula98,paredes98,taylor00,massi01,massi04}
made an interpretation in
terms of a collimated ejection with a constant position angle difficult.
The radio morphology  not only changes  position angle, but sometimes it is even one-sided while at other times
 two-sided. This fact suggested the hypothesis of \lsi being a precessing microblazar
 \citep{massi04}.

 A microblazar should be the galactic version of the extragalactic blazars, i.e., AGN with
 radio jets forming a small angle, $\theta$, with respect to the observer's line of sight.
 Doppler boosting enhances the radiation from material that is moving towards the observer,
  and attenuates it, if it moves in the opposite direction.
  For remarkable flux attenuation of the receding jet
  (i.e., attenuated to a level under the sensitivity of radio images) the structure will
  appear as a one-sided jet.
  A precession of the jet implies a variation of the  angle
  and therefore variable Doppler boosting. The result is both, a continuous
  variation of the position angle of the radio emitting structure
  and of the flux density  ratio between approaching and receding jet
 \citep{massi07}.

 In  known precessing X-ray binaries,
 the timescale for tidally forced precession of the accretion disk around the compact object, induced by the companion star, lies within the range
 $8-22$\,times the orbital period \citep{larwood98, massizimmermann10}.
 In this context, the peculiarity of the  variations of \lsi is their short timescale
  with respect to the predicted (8$-$22)$ \times 26.496$~d.
  In fact, MERLIN images revealed a surprising variation
  of 60\degr\, in position angle in only one day \citep{massi04}.
  Even if quantitatively the relationship between position angle in the image
  and the viewing angle $\theta$
  is not straightforward,  a fast  variation in position angle implies nevertheless clearly a fast variation in $\theta$.
  The fast position angle variation measured with MERLIN was  confirmed by VLBA observations
  \citep{dhawan06}  measuring  in  VLBA images a rotation of the inner structure of
  roughly 5\degr$-$7\degr\, in 2.5 hrs, that is again almost 60{\degr}/day.
  In that work some one-sided structures were taken as evidence for the cometary tail of the pulsar model.
  If the compact object is a pulsar,  the interaction between its   relativistic wind  and
  the   equatorial wind of the Be star  is predicted to create 
  a bow-shock around the pulsar
  with a sort of cometary tail, i.e., a one-sided structure, extending  away from the Be star 
   \citep{dubus06}.

However, recent  analysis of  the radio spectral index over 6.7 years from Green Bank Interferometer data at 2.2~GHz and 8.3~GHz discovered  new characteristics in the radio emission of \lsi  \cite{massikaufman09}.
%    However,  the  recent analysis of the radio spectral index by \citet{massikaufman09}
This analysis has proven that in \lsi occurs the typical characteristic of   microquasars
    of an optically thin outburst after  an interval of  optically thick emission.
    In microquasars,  the so called transient jet, associated to the large optically thin outburst,
    is related to shocks travelling in a pre-existing  steady jet,
    that is a slow moving continuous conical outflow with  a composite flat/inverted radio spectrum
    (i.e., optically thick emission) \citep{fender04, massi11}.
    The remarkable fact in \lsi is that,
    during the maximum of the long-term radio periodicity (4.6 yr) present in this source,
    the alternance, optically thick/thin emission,
    occurs twice  along the orbit,
    first around periastron and then again, shifted almost 0.3$-$0.4 in orbital phase towards  apastron
    (i.e, (0.3$-$0.4)$\times$ 26.496~d = (8$-$11)~d after periastron) \citep{massikaufman09}.
    This agrees  qualitatively and quantitatively with
     the well known ``two peak accretion/ejection model'',
     applied  by several authors in  \lsp,
predicting for large mass accretion rate,  $\dot{M}$, two events: one event around periastron and the second one 
shifted about 0.3 in orbital phase towards apastron
\citep{taylor92, martiparedes95, boschramon06, romero07}.

As a matter of fact, the radio spectral index data corroborate the microquasar model for \lsp.
Then, the open issue is what would produce the observed fast precession.
   % in a microquasar. 
   The basis for such a study is to establish the precessional period.
   This important parameter could be derived by the re-analysis of the VLBA observations performed 
   every 3 days over  30 consecutive days, towards the minimum of  the long-term radio periodicity
   \citep{dhawan06}.   We present here preliminary results of our analysis \citep{massi12}.
\section{Results}
\subsection{Images: the fast changing structure from a double-sided morphology to a one-sided morphology}
\begin{figure*}
 \centering
 \includegraphics[width=.75\textheight]{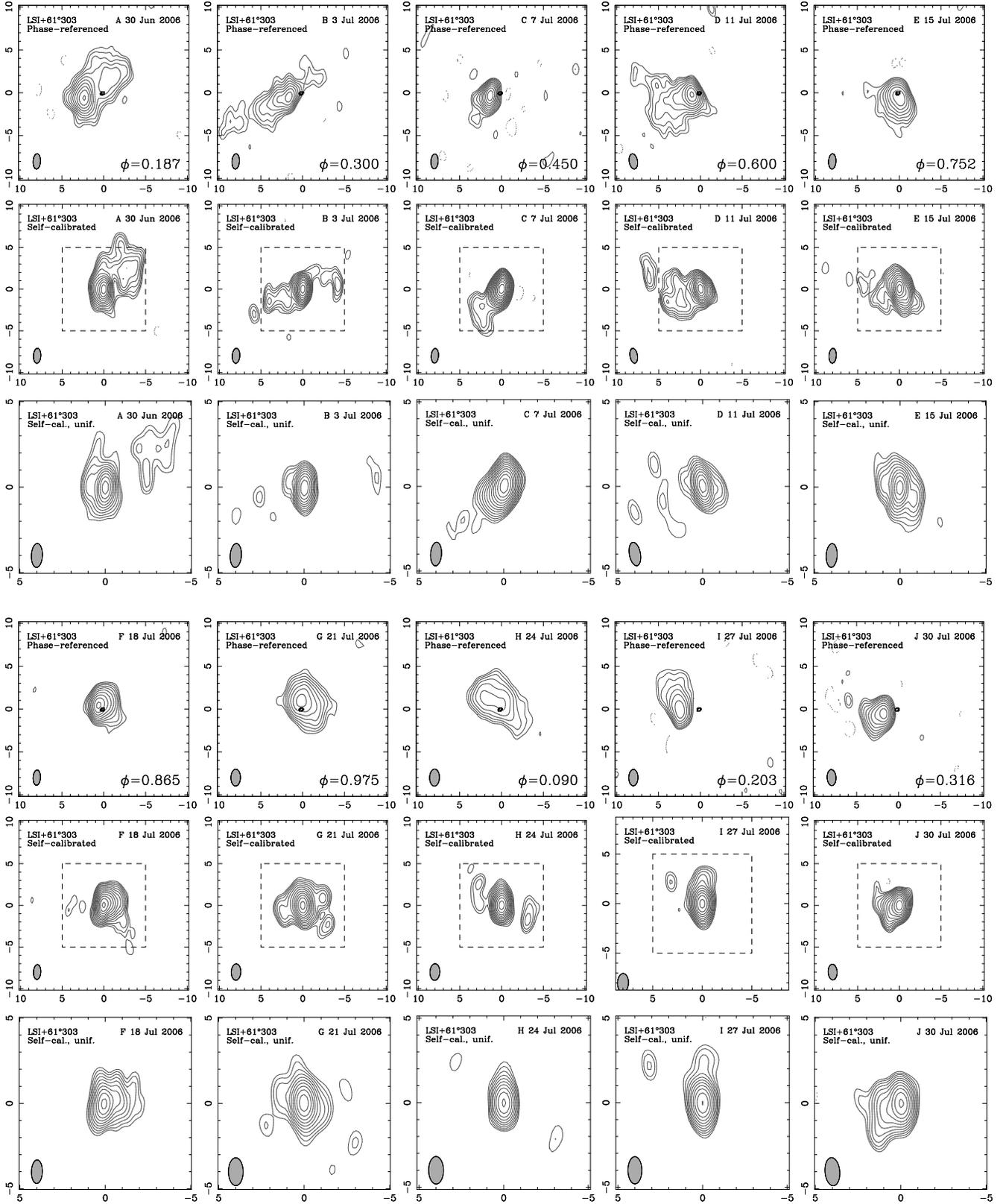}
       \caption{
       Images of VLBA  runs A-J at 8.4\,GHz (3.6\,cm) of \lsi \citep{massi12}.
       The units on the axes are milliarcseconds (mas).
       For each run
       three  maps  are presented, the phase-referenced map
        (beam of (1.8$-$2.0)  mas $\times$ (0.9$-$1.1)  mas, shown in the bottom left
	corner), the
	self-calibrated map at the same resolution and the self-calibrated  map
	with a beam of (1.4$-$1.6  mas $\times$  (0.6$-$0.9)   mas.
	Contour levels for all maps are $-$4, 4, 5.66, 8, 11.3, 16, 22.6, 32, 45.2,
	64, 90.5, 128, 181 $\sigma $
	(with $1 \sigma =0.1$ mJy/beam for Top and Bottom maps,  $1 \sigma \simeq 0.07$
	mJy/beam for Middle maps).
	%%%%%%%%%%%%%(with 1$\sigma$=0.1 mJy/beam for sel-calibrated maps). 
	At the center of the  phase-referenced images
	is traced, in scale,  the orbit (see Sect. 2).
	      }
	               \label{maps}
		          \end{figure*}

The 8.4~GHz  phase-referenced images
(with respect to  J0244$+$6228)
shown in Fig.~\ref{maps}-Top,
well reproduce the structures in \citep{dhawan06}.
At their center,  we show the orbit with semimajor axis of
$(0.22-0.23)$\,mas ($\simeq 0.7\times 10^{13}$ cm at a distance of 2\,kpc),
derived from Kepler's 3rd  law
for $P=26.496\pm 0.008$\,d with a mass for the  B0 star of 17\,$M_{\odot}$ and a mass of the compact object 
of $1.4 M_{\odot} - 4 M_{\odot}$, and traced using the 
orbital parameters from  \citep{casares05}.
The orbit  is  not only much smaller than the radio emitting region,
but also less than  the interferometric  beam of the VLBA at this wavelength.
We remark this, because  the zoomed orbit in  Fig.~3 of 
\citep{dhawan06} (zoomed to better show the orbital phase of each run)
may unfortunately produce  the wrong  impression that the orbit is resolved
    whereas it is not.
%, and only the accuracy of the astrometry, being 0.04\,mas at 8.4\,GHz, is less than the orbit.

	 %\subsection{8.4\,GHz  self-calibrated images}
	 For each run are presented, in Fig.~\ref{maps}, along with the  phase-referenced ones
	 also the  self-calibrated images with natural and uniform  weights
	 respectively.
	 The self-calibrated images were produced
	 with the automatic procedure Muppet  of the Caltech Difmap imaging package  \citep{shepard97}.
	 Using self-calibration, the information on absolute flux density and absolute position are lost,
	 but the removal of residual calibration errors improves the dynamic range
	 up to a factor of 4 for our maps.
%%%%% \citep{cornwellfomalont99}.
The higher dynamic range of the self-calibrated maps show
in the image of run A a double component structure.
The double structure in run A  is clearly unlikely for  a pulsar nebula,
whereas it could be compatible with either the approaching and receding component of a transient jet,
or with variable Doppler boosting along a one-sided twisted jet.
%(Zimmermann et al. in preparation).
Image B shows
already in the phase referenced map with natural weighting a double-sided morphology difficult to conciliate
with the pulsar nebula. The self-calibrated map shows
an intriguing  twisted jet as expected for a fast precession.
Runs C, F, I, and J show a one-sided jet or alternatively a pulsar cometary-tail but again
runs D, E, G,  and H show the double-sided jet structure.

\subsection{Astrometry: all points form one ellipse}

\begin{figure}
 \centering
 \includegraphics[width=.47\textwidth, clip, angle=0]{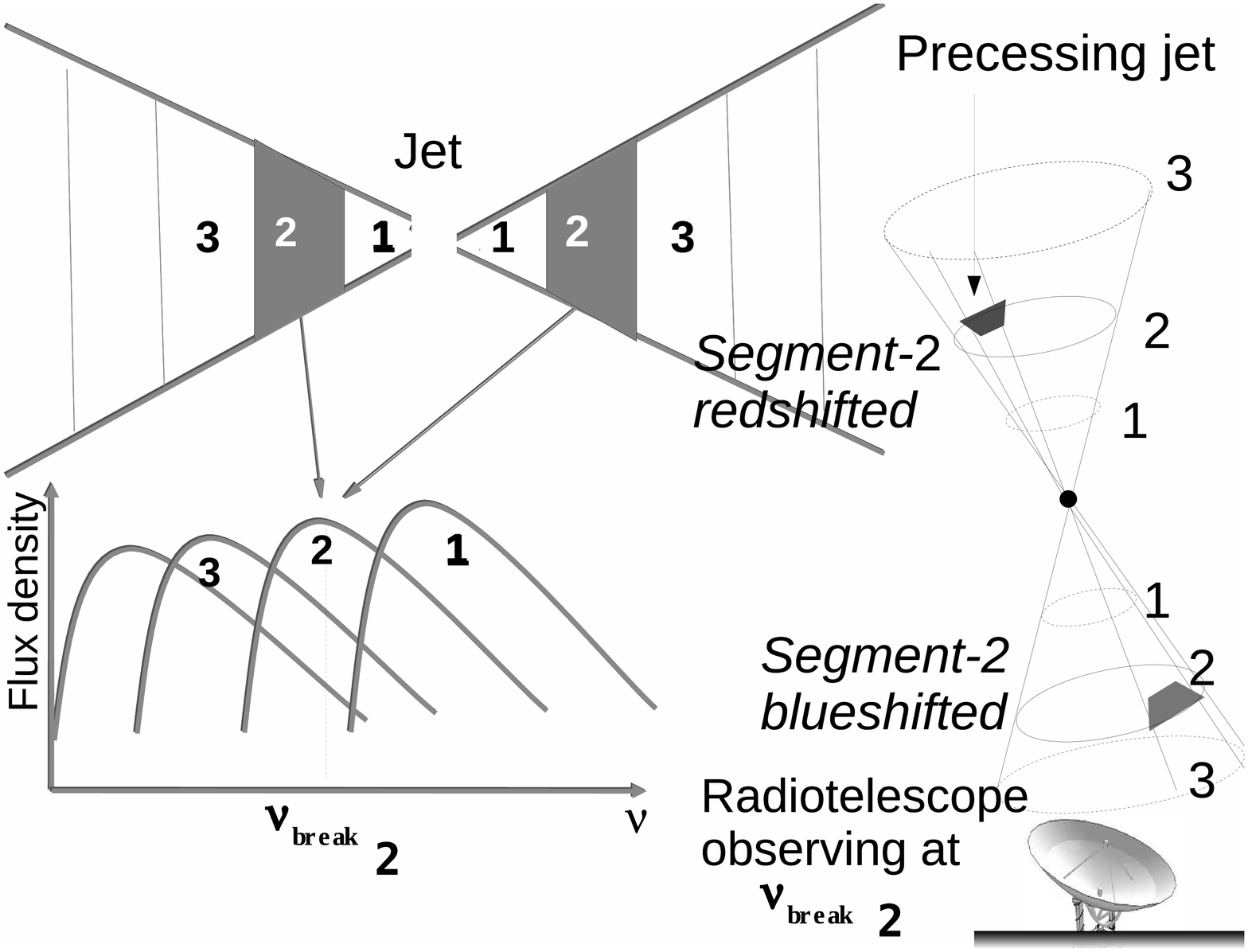}\\
 \includegraphics[width=.35\textwidth, clip, angle=-90]{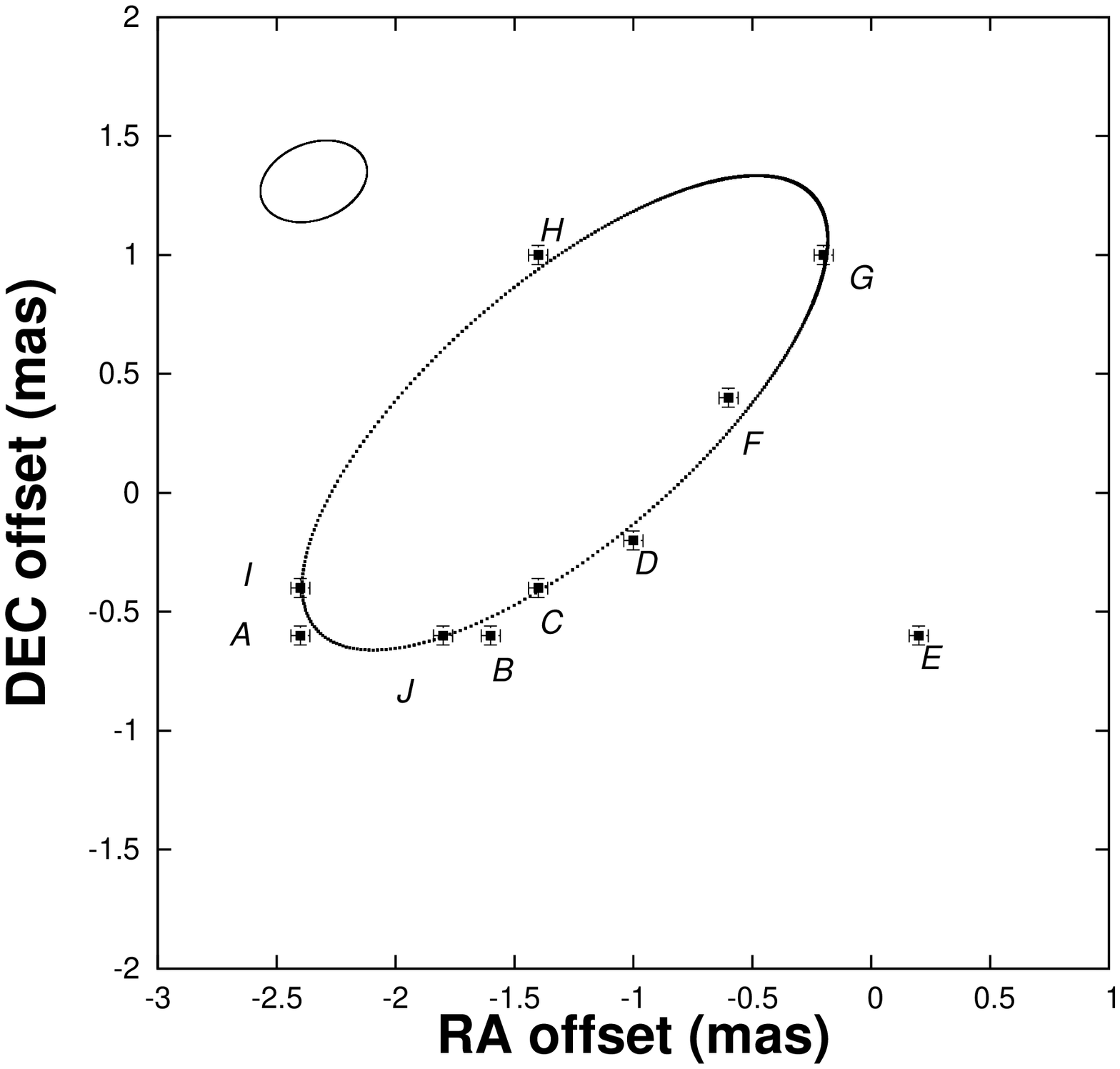}\\
 \caption{
Top-Left: Superposition of individual spectra,  each with a different break frequency associated with different  segments of a steady jet. Top-Right: For a precessing microblazar, i.e., for small angle of ejection with respect to the line of sight, the  core component,  dominated by the approaching jet contribution because of Doppler boosting, will describe an ellipse during  precession. Bottom: Astrometry of consecutive peaks of  VLBA  8.4 GHz maps for runs A-J with the orbit drawn in scale at an arbitrary distance. The quality of the astrometry is determined
from the scatter in positions of the check source 
 J0239+6005 \citep{massi12}. The peak of map E, even if displaced from the other ones,  and therefore likely affected by the approaching component of the transient jet, is at a position angle which is consistent  with  the other peaks. The fact that the precessing compact object is moving in an orbit with semimajor axis of $\sim$0.2 mas introduces additional variations \citep{massi12}.
 }
 \label{sketch}
 \end{figure}

The peak position in the phase-referenced maps are plotted  in  Fig.~\ref{sketch}-Bottom.
It is evident that the  peaks trace well an ellipse
6$-$7  times larger than the  orbit.
Our hypothesis is that this ellipse is the cross-section of the
precession cone, at the distance
of the 3.6 cm-core  ($\nu = 8.4$~GHz) of the steady jet.

In fact, changes in the electron energy distribution and the
decay of the magnetic field
along the  conical outflow forming the steady jet,  imply that the critical  frequency, $\nu_\mathrm{break}$, 
of the synchrotron spectrum  varies along the jet \citep{marscher95, massi11}.
In microquasars, the  $\nu_\mathrm{break}$ for the part of the steady jet closest to the engine (i.e.,
 $\nu_\mathrm{break_1}$ in Fig.~2) lies in the infrared \citep{russell06},
        whereas in AGN the observed turnover
	%is at $10^{11\pm 0.5}$ Hz, called ``millimeter-wave core'' in the literature
	lies in the millimeter range \citep{marscher95}.
	Observing at longer wavelengths than infrared, i.e., in the  radio band, gives
	rise to two results: a flat spectrum and the ``core shift''.
	Multi-wavelength observations result in fact in the flat/inverted spectrum discussed in the introduction,
	typical for a steady jet.
	Imaging a steady jet at one observing frequency  $\nu_\mathrm{obs}$,
	gives rise to the effect known as ``core shift''\citep{lobanov98}, with the displacement  from the center  as a
	function of the observing frequency  $\nu_\mathrm{obs}$.
	At $\nu_\mathrm{obs}$ the emission of the segment will dominate, whose spectrum peaks at
	that frequency
	plus small contributions from neighboring segments (see Fig.~1 in \citep{markoff10}).
	With the engine being  close to the ``infrared-core", it is clear that
	the 3.6~cm-core can be rather displaced from the center (like  segment 2 is displaced with respect to
the center in  Fig.~2 Top-Left) \citep{marscher95, massi11, lobanov98}.
	If a jet  is pointing  towards us (i.e., a microblazar as in Fig.~2 Top-Right)
	the core will be dominated by
	the approaching jet,  because of Doppler boosting.
	If the jet is precessing, then the core will  describe
	an  ellipse (see both ellipses, the observed ellipse in Fig.~2-Bottom and the 
predicted ellipse for the blueshifted segment-2 in  
      Fig.~2 Top-Right).
	A transient jet, on the contrary, can have any shift from the center, depending on the velocity
	and the  time elapsed from the transient.
	   %In Fig.~3, we see that the consecutive peaks trace well an ellipse. 
	   The peak of map E, even if displaced from the other ones, and therefore likely affected by 
the approaching component of the transient jet, is at a position angle which is consistent  with  the other peaks.
%The precession period relative to the traced ellipse is about 28~d
%(by comparing the position of  I (27 July) and J(30 July) with that of  A
%(30 June)).   
%

Note as  after 27~d, the peak of run I is  rather close to completing the  cycle, i.e. to
 overlap  with the peak of the starting run A. The same occurs for peak of run J, that 27 d after run B,
 nearly overlaps  with peak B.
 The peak of run J, 30 d after run A, is clearly rather
 displaced from the A peak. The period seems therefore to be in the range (27$-$28) d.

\section{Conclusions and Discussion}
\lsi is one of the few established massive X-ray binaries that emit in the high and very high energy range. It is formed by a compact object of unknown nature (black hole or neutron star) travelling with a period of 26.496 d around a Be star. Two models have been proposed in the past: a millisecond pulsar and a microquasar model. Recent analysis of the radio spectral index  has proven in \lsi the typical characteristic of microquasars of an optically thin outburst after an interval of optically thick emission twice along the orbit, as predicted by the two peak microquasar model.

The results of
our re-analysis of VLBA observations of \lsi presented here are that  the radio emission has in several images a double-sided structure (see. Fig. 1). The astrometry show that the peaks of   the 
images trace a well defined  ellipse in (27$-$28) d (see Fig. 2). 
The  pulsar model explains neither the double-sided morphology nor the observed change from double sided to a  one-sided structure. The microquasar model can explain them with variable Doppler boosting, i.e., with a  precessing jet. The cm-core of a  precessing steady jet pointing close to our line of sight, as in a microblazar, is  expected to describe an ellipse during the precession. During the transient jet phase there will be an additional shift due to the approaching jet component. We conclude therefore that the precession period is the time of (27$-$28)~d necessary to complete the ellipse. 

A precession period  of (27$-$28) d for the accretion disk in \lsi
induced by tidal forces of the Be star \citep{massizimmermann10} 
would require the unrealistic value for the size of the accretion disk of $0.5-0.8  \times 10^{13}$~cm, i.e., nearly the semi-major axis, 
and can therefore be ruled out. On the contrary, for a slow rotating compact object, a precession period of 28 d induced by Lense-Thirring effect 
(i.e., frame dragging produced by the rotation of the compact object)
could be compatible \cite{massizimmermann10}.

High energy observations of \lsi are consistent with the microblazar nature of this object.
Simultaneous X-ray and VHE observations during an outburst 
of \lsi resulted in a 
correlation,  indicating 
a simultaneity in the emission processes
\citep{Anderhub09}.
In particular, with respect to the comparison with blazars, it has been noticed  \citep{massizimmermann10}  
as the two fluxes result in $F_{\rm VHE}\propto F_{\rm X}^{\eta}$ with $\eta=0.99$ 
in agreement with the correlation observed in blazars \citep{Katarzynski2010}. 
%%%%%%Besides the X-ray/VHE emission
%correlation, 
In addition, the respective values of the  photon indices seem to be comparable.
%Concerning the photon index of such emission, 
%\citet{Acciari2011} report that 
During a bright flaring event of the blazar 1ES 2344+514, VERITAS measured a photon index of $\Gamma=2.43 \pm$ 0.22,
whereas for the X-ray emission, RXTE and Swift-XRT measured a hard photon index of $\sim$1.9
\citep{Acciari2011}. 
%%%%%%%%%This is comparable with the photon indices 
%measured by \citet{Anderhub09} 
For \lsp: MAGIC had a $\Gamma=2.7$ 
and the X-ray emission detected by \textit{XMM-Newton} and Swift-XRT a harder photon index of 1.5-1.8 \citep{Anderhub09}.
In blazars the X-ray emission is due to synchrotron, and VHE is synchrotron self-Compton (SSC) \citep{Katarzynski2005}. 
As discussed in \citep{zimmermannmassi12}, several authors have in fact explained the X-ray excess around apastron in \lsi with synchrotron and the VHE with either external inverse Compton (EIC) or SSC emission \citep{Gupta2006,Zabalza2011}. 
%\citet{Zabalza2011} show that 
In particular, the X-ray/VHE correlation in \lsi  \citep{Anderhub09} is compatible with a one zone leptonic particle population producing the emission 
\citep{Anderhub09,Zabalza2011}.
%%%%%%%%%%The model, that uses the observed X-ray photon index (1.55-1.67),   reproduces well the lightcurves \citep{Zabalza2011}. 
Concerning high energy emission in the GeV range, detected e.g., with \textit{Fermi}-LAT, 
GeV emission is seen all along the orbit. In fact, electrons from the steady jet can always upscatter stellar UV photons to GeV energies
(i.e., EIC see \citep{boschramon06}). 
Nevertheless, more energetic particles from the transient jet could in addition also produce GeV emission via EIC and SSC. Intriguingly, the spectrum measured by \textit{Fermi}-LAT shows, in addition to a power law with a cut-off around 6 GeV, upper limits possibly compatible with the spectrum measured with MAGIC and VERITAS (see e.g., Figs. 2 and 3 in \citep{Hadasch2010} and discussion in \citep{zimmermannmassi12}).
As a matter of fact, there is  an interesting increase in the overall flux level observed with \textit{Fermi}-LAT after March 2009 together with a broadening of the peak shape \citep{Hadasch2010}. The up to now observed  \textit{Fermi}-LAT variations are therefore consistent with 
a long term variation. Similarly, strong variations are observed at very high energies. With VERITAS the source went from being detected around apastron 
to becoming quiescent between 2008 and 2010 \citep{acciari11}.
With respect to the long-term,  4.6 yr, radio periodicity mentioned in Sect. 1,  
the insufficient temporal  coverage at high energy, evident in Fig. 1 of \citep{zimmermannmassi12}, does not allow at the moment a more closer comparison. However, a different trend with respect to the  4.6 yr radio periodicity is expected. 
In fact, the high energy emission is indeed related to the steady jet but its peak  seems to occur during the transient jet
and not during the steady one (Fig. 4 in \citep{massi11}). As a matter of fact 
the steady jet, related to the radio periodicity,  is not always followed by the transient one.
A timing  analysis  of the transient jet  is in progress.


\begin{thebibliography}{}
\bibliographystyle{apsrev}

  \bibitem[Massi 
  \& Kaufman Bernad{\'o}(2009)]{massikaufman09} Massi, M., \& Kaufman Bernad{\'o}, M.,
``Radio Spectral Index Analysis and Classes of Ejection in \lsp'',
\ 2009, ApJ, 702, 1179 

  \bibitem[Massi et al.(1993)]{massi93}
  Massi, M., Paredes, J.M., Estalella, R., \& Felli, M.,
``High resolution radio map of the X-ray binary \lsp'',
  1993, A\&A, 269, 249

  \bibitem[Massi et al.(2001)]{massi01} Massi, M., Rib{\'o}, M., Paredes, J.~M., Peracaula, M., \& Estalella, R.,
``One-Sided Elongated Feature in \lsp'',
\ 2001, A\&A, 376, 217 

  \bibitem[{Massi et al.(2004)}]{massi04} Massi, M., Rib\'o, M,  Paredes, J. M., et al.,
``Hints for a fast precessing relativistic radio jet in \lsp'',
 2004, A\&A, 414, L1

  \bibitem[Peracaula et al.(1998)]{peracaula98} Peracaula, M., Gabuzda, D.~C., \& Taylor, A.~R.,
``Rapid expansion in the VLBI structure of \lsp'',
\ 1998, A\&A, 330, 612 

  \bibitem[Paredes et al.(1998)]{paredes98} Paredes, J.~M., Massi, M., Estalella, R., \& Peracaula, M.,
``Milliarcsecond radio structure of \lsp'', \ 1998, A\&A, 
  335, 539 

   \bibitem[Taylor et al.(2000)]{taylor00}
   Taylor, A.R., Dougherty, S.M., Scott, W.K., Peracaula, M., \& Paredes, J.M.,
``VSOP Imaging of the Unusual X-Ray Binary Star'',
   2000, Proc. of  Astrophysical Phenomena Revealed by Space VLBI, eds.
   H. Hirabayashi, P.G. Edwards, and D.W. Murphy., p.~223


  \bibitem[Massi(2007)]{massi07} Massi, M.,
``The Enigmatic Compact Object in the Stellar System \lsp: Accreting or Not Accreting?'',
\ 2007, The
  Multicolored Landscape of Compact Objects and Their Explosive Origins, 924,
  729

  \bibitem[Larwood(1998)]{larwood98} Larwood, J.,
``On the precession of accretion discs in X-ray binaries'',
\ 1998, MNRAS, 299, L32

  \bibitem[Massi \& Zimmermann(2010)]{massizimmermann10} Massi, M., \&
  Zimmermann, L., ``Feasibility study of Lense-Thirring precession in \lsp'',  \ 2010, A\&A, 515, A82

 \bibitem[Dhawan et al.(2006)]{dhawan06} Dhawan, V.,  Mioduszewski, A., \&
 Rupen, M.,
``\lsi is a Be-Pulsar binary, not a Microquasar'',
 2006, Proceedings of  the VI Microquasar Workshop, p. 52.1

 \bibitem[Dubus(2006)]{dubus06} Dubus, G.,
``Gamma-ray binaries: pulsars in disguise?'',
 \ 2006, A\&A, 456, 801

 \bibitem[Fender et al.(2004)]{fender04}
  Fender, R.~P., Belloni, T.~M., \& Gallo, E.,
``Towards a unified model for black hole X-ray binary jets'',
\ 2004, MNRAS, 355, 1105


  \bibitem[Massi(2011)]{massi11} Massi, M.\,
``Steady jets and transient jets: observational characteristics and models'',
 2011,  Mem. Soc. Astron. It., 82, 24

 \bibitem{boschramon06} Bosch-Ramon, V., Paredes,
 J.~M., Romero, G.~E., \& Rib{\'o}, M.,
``The radio to TeV orbital variability of the microquasar \lsp'',
\ 2006, A\&A, 459, L25

  \bibitem[Marti \& Paredes(1995)]{martiparedes95} Marti, J., \& Paredes,
  J.~M.,
``Modelling of \lsi from near infrared data'',
\ 1995, A\&A, 298, 151

  \bibitem[Romero et al.(2007)]{romero07} Romero, G.~E., Okazaki, A.~T., Orellana, M., \& Owocki, S.~P.,
``Accretion vs. colliding wind models for the gamma-ray binary \lsp: an assessment'',
\ 2007, A\&A, 474, 15 

   \bibitem[Taylor et al.(1992)]{taylor92} Taylor, A.~R., Kenny, H.~T.,
   Spencer, R.~E., \& Tzioumis, A.,
``VLBI observations of the X-ray binary \lsp'',
 1992, ApJ, 395, 268


  \bibitem[Massi et al.(2012)]{massi12} Massi, M., Ros, E.,  
\&  Zimmermann, L.,
``VLBA images of the precessing jet of \lsp'',
  2012  A\&A, 540, A142 

 \bibitem[Casares et al.(2005)]{casares05} Casares, J., Ribas, I., Paredes,
 J.~M., Mart{\'{\i}}, J., \& Allende Prieto, C.,
``Orbital parameters of the microquasar \lsp'',
\ 2005, MNRAS, 360, 1105

  \bibitem[Shepherd(1997)]{shepard97}
  Shepherd, M. C.,
  ``Difmap: an Interactive Program for Synthesis Imaging'',
  1997, Astronomical Data Analysis Software and Systems VI, A.S.P. Conference Series
  Gareth Hunt and H. E. Payne, eds.,   125, 77

  \bibitem[Marscher(1995)]{marscher95} Marscher, A.~P.,
`Probes of the Inner Jets of Blazars'', \ 1995,
  Proceedings of the National Academy of Science, 92, 11439

  \bibitem[Russell et al.(2006)]{russell06}
  Russell, D.~M., Fender, R.~P., Hynes, R.~I., Brocksopp, C.,
   Homan, J., Jonker, P.~G.,
   \& Buxton, M.~M.,
``Global optical/infrared-X-ray correlations in X-ray binaries: quantifying disc and jet contributions'',
\ 2006, MNRAS, 371, 1334

  \bibitem[Markoff(2010)]{markoff10} Markoff, S.,
``From Multiwavelength to Mass Scaling: Accretion and Ejection in Microquasars and AGN'',
\ 2010, Lecture
  Notes in Physics, Berlin Springer Verlag, 794, p.143

\bibitem[Lobanov(1998)]{lobanov98} Lobanov, A.~P.,
``Ultracompact jets in active galactic nuclei'',
\ 1998, A\&A, 330, 79 

  \bibitem[Anderhub \& al.(2009)]{Anderhub09} Anderhub, H., et al.,
``Correlated X-Ray and Very High Energy Emission in the Gamma-Ray Binary \lsp'',
\ 2009,
  ApJ, 706, L27

\bibitem[Katarzy\`nski \& Walczewska(2010)]{Katarzynski2010} Katarzy\`nski, K., Walczewska, K.,
``On the correlation between the X-ray and gamma-ray emission in TeV blazars'',
 \ 2010, A\&A, 510, A63


%\bibitem[Acciari et al.(2011b)]{Acciari2011b} Acciari, V.~A., et al.\ 
%2011b, arXiv:1106.4594 
\bibitem[Acciari et al.(2011)]{Acciari2011} Acciari, V.~A., Aliu, 
E., Arlen, T., et al.,
``Multiwavelength Observations of the Very High Energy Blazar 1ES 2344+514'',
\ 2011, ApJ, 738, 169 

\bibitem[Katarzy{\'n}ski et al.(2005)]{Katarzynski2005} Katarzy{\'n}ski, K., et al.,
``Correlation between the TeV and X-ray emission in high-energy peaked BL Lac objects'',
 2005, A\&A, 433, 479

  \bibitem[Zimmermann \& Massi(1012)]{zimmermannmassi12} Zimmermann, L. \& Massi, M.,
``Implications of the radio spectral index transition in \lsi for its INTEGRAL data analysis'',  
  2012, A\&A, 537, A82 

\bibitem[Gupta \& Boettcher(2006)]{Gupta2006} Gupta, S. \& Boettcher, M.,
``A Time-dependent Leptonic Model for Microquasar Jets: Application to \lsp'',
\ 2006, ApJ, 650, L123

\bibitem[Zabalza et
al.(2011)]{Zabalza2011} Zabalza, V., Paredes, J.~M., \& Bosch-Ramon, V.,
``On the origin of correlated X-ray/VHE emission from \lsp'',
\ 2011, A\&A, 527, A9

\bibitem[Hadasch(2011)]{Hadasch2010} Hadasch, D., \& for the Fermi-LAT collaboration,
``Results from the binaries \lsi  and LS 5039 after 2.5 years of Fermi monitoring'',
Fermi Symposium proceedings-eConf C110509, ed. Morselli, 2011, [{\tt arXiv:1111.0350}]

%%%%%%%%%%\bibitem[Albert et al.(2009)]{Albert09} Albert, J., et al.\ 2009, ApJ, 693, 303
\bibitem[Acciari et al.(2011)]{acciari11} Acciari, V.~A., Aliu, 
E., Arlen, T., et al.,
``VERITAS Observations of the TeV Binary \lsi During 2008-2010'',
\ 2011, ApJ, 738, 3 


\bibitem[Massi(2011)]{massi11} Massi, M.,
``The two-peak model of LS I +61303: radio spectral index analysis'',
\ 2011, Mem. Soc. Astron. It., 82, 77 

  %\bibitem[Marscher \& Gear(1985)]{marschergear85}
  %Marscher, A.~P., \& Gear, W.~K.\ 1985, \apj, 298, 114

 % \bibitem[Mart\'\i-Vidal \& Marcaide(2008)]{martividal08}
 % %Marti-Vidal, I.  \& J. M. Marcaide, J. M. 2008,
 % Mart\'\i-Vidal, I.  \& J. M. Marcaide, J. M. 2008,
 % A\&A, 480,   289



  %\bibitem[Miller-Jones(2008)]{millerjones08} Miller-Jones, J.~C.~A.\ 
  %2008, Journal of Physics Conference Series, 131, 012057 

%  \bibitem[Mirabel 
%  \& Rodr{\'{\i}}guez(1999)]{mirabelrodriguez99} Mirabel, I.~F., \& Rodr{\'{\i}}guez, L.~F.\ 1999, ARA\&A, 37, 409 

  %\bibitem[Mold{\'o}n et al.(2011)]{moldon11} Mold{\'o}n, J., 
  %Johnston, S., Rib{\'o}, M., Paredes, J.~M., 
  %\& Deller, A.~T.\ 2011, \apjl, 732, L10 


 % \bibitem[Pradel et 
 % al.(2006)]{pradel06} Pradel, N., Charlot, P., \& Lestrade, J.-F.\ 2006, A\&A, 452, 1099 

 % \bibitem[Rickett(1990)]{rickett90} Rickett, B.~J.\ 1990, ARA\&A, 28, 561 


%\begin{thebibliogArapAhy}{99} % Use for 10-99 references

%\bibitem{accelconf-ref}
%http://www.cern.ch/accelconf

%\bibitem{exampl-ref}
%A.N. Other, ``A Very Interesting APapAer'', EPAC'96, Sitges, June
%1996.

%\bibitem{templates-ref}
%http://www.cern.ch/accelconf/templates.html

\end{thebibliography}
%\begin{thebibliography}{9}   % Use for  1-9  references
%\bibliographystyle{plainnat}
%\bibliographystyle{aa}

\end{document}